\documentclass{llncs}

\usepackage{amsmath}
\usepackage{amsfonts}
\usepackage{graphicx}
\usepackage{subcaption}
\usepackage{algorithmic,algorithm}
\usepackage{color}
\begin{document}
\title{Handling oversampling in dynamic networks using link prediction\thanks{This
work is sponsored by the Assistant Secretary of Defense for Research \&
Engineering under Air Force Contract FA8721-05-C-0002.  Opinions,
interpretations, conclusions and recommendations are those of the authors and
are not necessarily endorsed by the United States Government.}}

\author{Benjamin Fish\inst{1,2} and Rajmonda S. Caceres\inst{2}}
\institute{University of Illinois at Chicago, Chicago, IL \and
MIT Lincoln Laboratory, Lexington, MA\\
}
\maketitle

\begin{abstract}Oversampling is a common characteristic of data representing dynamic networks.  It introduces noise into representations of dynamic networks, but there has been little work so far to compensate for it.  Oversampling can affect the quality of many important algorithmic problems on dynamic networks, including link prediction.  Link prediction seeks to predict edges that will be added to the network given previous snapshots.  We show that not only does oversampling affect the quality of link prediction, but that we can use link prediction to recover from the effects of oversampling.  We also introduce a novel generative model of noise in dynamic networks that represents oversampling.  We demonstrate the results of our approach on both synthetic and real-world data.
\end{abstract}

\section{Introduction}
Networks have become an indispensable data abstraction that captures the nature of a diverse list of complex systems, such as online social interactions and protein interactions. All these systems are inherently dynamic and change over time. A common abstraction for incorporating time has been the ``dynamic network," a time series of graphs, each graph representing an aggregation of a discrete time interval of the observed interactions. While in many cases the system under observation naturally suggests the size of such a time interval, it is more often the case that the aggregation is arbitrary and is done for the convenience of the data representation and analysis. However, an abundance of literature has demonstrated that the choice of the time interval at which the network is aggregated has great implications on the structures observed and inferences made~\cite{clauset2012persistence,sulo2010meaningful,holme2012temporal,ribeiro2013quantifying}.

\subsection{Oversampling}

We view the system through the filter of the data we collect. This data is typically collected opportunistically, with the temporal rate of data not always matching that of
the system.  With the advent of microelectronic data collection systems such as GPS and RFID sensors, it is often the case that data is sampled at orders of magnitude more frequently than the
temporal scale of the underlying system. Therefore, it is important that the aggregation process that transforms the collected data into a dynamic network representation correctly accounts for the oversampling effects. 

Oversampling is an aspect of the data collection process that can help with the issue of representing continuous time discretely. It helps reduce the number of missing interactions and allows us to better identify persistent interactions. On the other hand, oversampling affects our ability to distinguish between noisy local temporal orderings and critical temporal orderings. For example, when analyzing email communication networks, the data is collected at a resolution of seconds, but the causality of email interactions and the emergence of complex structures such as communities is often more accurately represented and detectable at much coarser scales.


The concept of oversampling has been studied extensively in the signal processing community.  However, to the best of our knowledge, there is no natural, well-defined notion of oversampling that translates directly to the domain of graph sequences.  So here we use oversampling to mean a {\it distribution over graph sequences} that displays the typical effects of sampling too frequently:  noisy local temporal orderings, spurious interactions, etc.

In this paper, we assume we are given a noisy dynamic network that has been observed at a fixed oversampled rate.  Our goal is to recover from this oversampling by aggregating the network together in such a way as to remove any artifactual temporal orderings, while preserving any real temporal information such as edge co-occurrences and critical temporal orderings. We represent a dynamic network as a sequence of discrete snapshots, that is, as a sequence of graphs.  Given a window size $w$, we bin together into a graph all edges that occur within each length-$w$ time span. This results in a new dynamic network. The goal becomes to find the window size $w$ that best recovers from the noise.  At one extreme, if this window size $w$ is the entire length of the original dynamic network, then this means that there is no temporal information and the network is static. At the other extreme, if this window size $w$ lasts for a single snapshot of the original dynamic network, then this means there is no oversampling at all and the network is observed at the right temporal scale.  Thus, the window size can be seen as a proxy for the amount of temporal information stored in the dynamic network, and finding a good window size can be seen as finding the temporal scale at which the network is evolving.

\subsection{Link Prediction}
In this work, we take the approach that the inference task should inform how to recover from oversampling, whether the task is link prediction, community detection, or something else. Link prediction, in particular, is an important inference task with many applications.
It has beeen used in the analysis of the internet~\cite{adafre2005discovering,zhu2002using}, social networks~\cite{adamic2003friends}, and biological networks~\cite{airoldi2006mixed,freschi2009graph}.  It has also been used for designing recommendation systems~\cite{huang2005link,liu2007predicting} and classification systems~\cite{gallagher2008using}.  See the survey of Al Hasan and Zaki~\cite{al2011survey} for more applications and an introduction to the various techniques used in link prediction.

Given the importance of link prediction, we investigate whether this task can serve as a good driver for recovering the correct temporal scale of an oversampled dynamic network. Our approach is simple:  we use the quality of a link prediction algorithm as a score for the window size. If we can predict links better, it means that we have found a good window size, not only because by definition we have improved the performance of our inference task, but because at this scale we can better capture the evolution of the network.  

To our knowledge, we are the first to formally study the relationship between temporal oversampling in networks and link prediction.  As such, we need to introduce a model of noise for dynamic networks that captures effects of oversampling. We define such a model by leveraging the following observation: when a dynamic network is oversampled, edge occurrences are recorded  near the ``natural'' occurrence time, but are spread out around that time. We present a model that generates noisy, oversampled dynamic networks by distributing the times of edges over a Gaussian distribution.  This model can be extended to capture unique characteristics of a given data collection process. It also is applied independently of the underlying process generating the network, allowing for a wide variety of phenomena to be modeled.

In this paper, we show that oversampling affects the quality of link prediction algorithms. Furthermore, using synthetic data, we show that link prediction performs better on graph sequences aggregated near the ground-truth window size than on other window sizes.  In this light, we can recover from the effects of oversampling by using link prediction to find a good window size.  Finally, we show that our method results in robust results on real-world networks.

In Section~\ref{sec:defs}, we give a more formal description of the oversampling problem for dynamic networks and present our link prediction based approach. We define a generative model for Gaussian-type noise representing oversampling in Section~\ref{sec:models}, as well as a generative model for dynamic networks for testing link prediction algorithms.  In Section~\ref{sec:gaussiannoise}, we use our synthetic generative model to test our approach.  We demonstrate that our method yields reasonable results and show the impact of a variety of different parameters on our results.  We then show the results of our approach on two real-world networks in Section~\ref{sec:real data}.  We end with a few concluding remarks in Section~\ref{sec:conclusion}.

\subsection{Related Work}\label{subsec:prev work}

Extensive literature has demonstrated that the choice of aggregation window greatly impacts the quality of the corresponding dynamic network~\cite{clauset2012persistence,ribeiro2013quantifying,sulo2010meaningful}.
Some work has been done in developing heuristics for identifying the ``right" window size or temporal partitioning, especially for numerical time series ~\cite{6137319,Wagner:2008}.  In dynamic networks, though, most work does so only in limited or slightly different contexts.  Peel and Clauset, for example, consider the problem of finding change points, points at which the generative process of a dynamic network is itself changing~\cite{DBLP:journals/corr/PeelC14}.  Sun et al.~also considers the problem of finding change points - and more generally considers the problem of finding a partition of the network - this time in the context of community detection~\cite{sun2007graphscope}. 
In~\cite{eagle2006reality} and~\cite{clauset2012persistence}, they analyze the discrete Fourier transform of time series of different graph metrics to identify important frequencies in a dynamic network.
Similarly, Sulo et al.~\cite{sulo2010meaningful} use information-theoretic tools to analyze time series of a variety of different graph metrics for graph sequences that have been aggregated at different temporal scales.

Prediction as a tool to inform model selection is not new, especially in the literature of time series analysis.  
Central in this literature is the principle of minimum description length, which says that if data displays any regularity (i.e. is predictable), then that regularity can be used to shorten an encoding of the data.  Therefore finding a short encoding is finding a good predictor, and vice versa.  See~\cite{hansen2001model} for a survey on the subject.  In the context of temporal networks, this approach is not as common, but link prediction has been used to infer useful static networks from data \cite{de2010inferring}.

 There are a number of models in the literature for dynamic networks, including the GHRG model of Clauset and Peel~\cite{DBLP:journals/corr/PeelC14}, the activity-driven model of Perra et al.~\cite{perra2012activity}, and the dynamic latent-space model of Sarkar and Moore~\cite{sarkar2005dynamic}.  Some models for static networks can also naturally be seen as a model for dynamic networks in which the number of nodes is growing over time, such as in the preferential attachment model.  However, none of these models are well-suited for modeling oversampling or for controlling the quality of a link prediction algorithm as we will need, so we use a novel model, which we detail in Section~\ref{sec:models}.

\section{Problem Formulation and Methodology}\label{sec:defs}

\begin{figure}[h]
\begin{center}
\includegraphics[scale = 0.25]{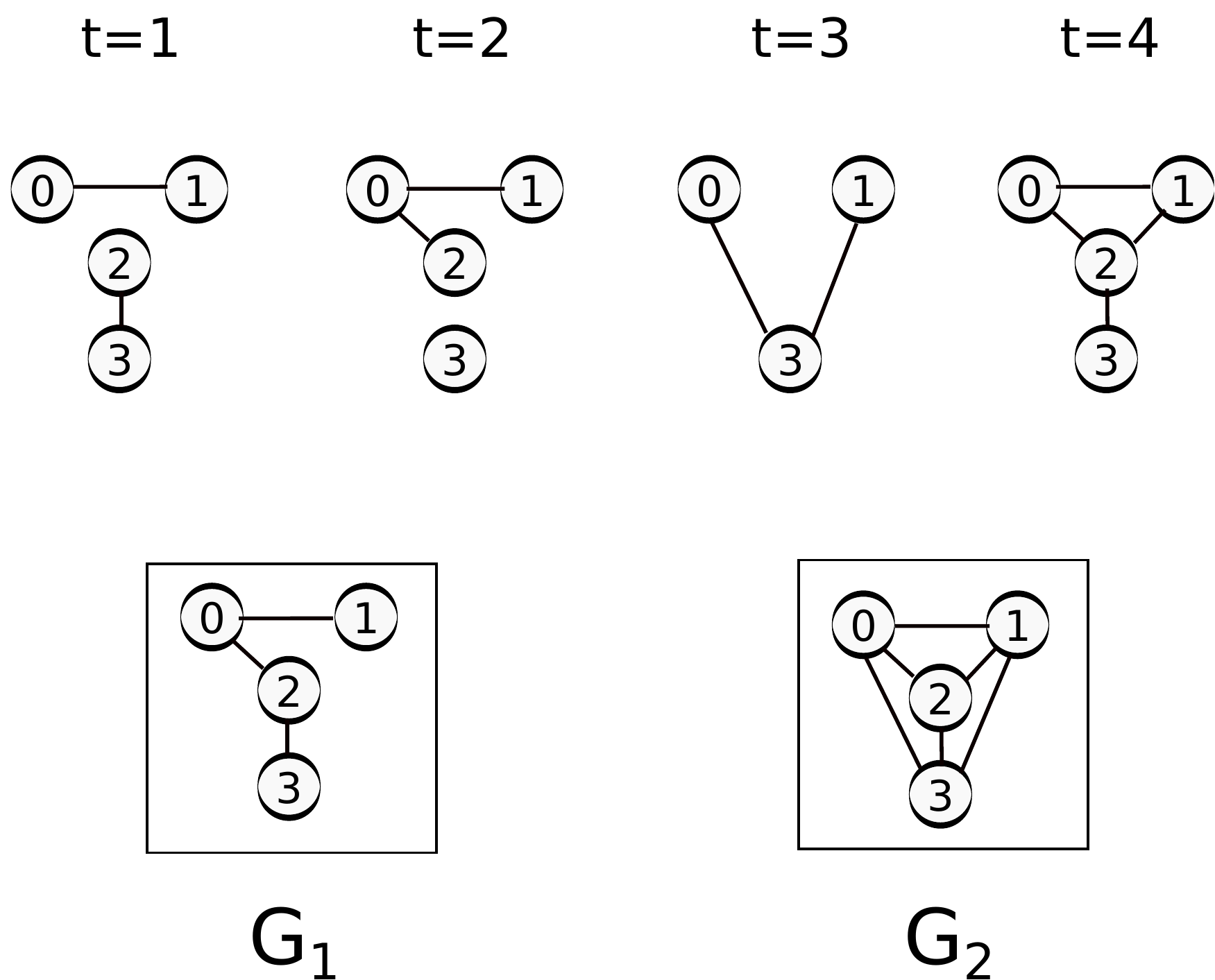}
\end{center}
\caption{Example of aggregating a time series of graphs into a coarser window $w=2$.}
\label{fig:dynNet}
\end{figure}

To formalize the problem, we assume that a noisy dynamic network is a sequence of discrete graphs (each graph representing one time step) on a fixed set of $n$ nodes.  Furthermore, this graph sequence is the output of some noise process being applied to a (possibly shorter) sequence of graphs - the `ground truth' - on the same set of $n$ nodes.  Specifically, we assume that the noisy sequence can be partitioned into `windows' of fixed size, and that any temporal orderings between edges in a single window are spurious.  Thus recovery consists of the two step process of first choosing a window size $w$ (which in this paper we will also refer to as a \emph{temporal scale} of the graph sequence) and then aggregating all edges within each $w$ consecutive graphs into a single graph. This results in a new graph sequence that is a factor of $w$ shorter.  Figure~\ref{fig:dynNet} gives an illustration of this process.

In general, the goal is to recover the ground truth from the noisy sequence.  Of course, this is not always possible.  In the degenerate case, the ground-truth sequence is just noise and this task is impossible. In this paper, we assume the ground-truth sequence is sufficiently non-noisy, so that a link prediction algorithm will perform well on it.  Link prediction is the classification task of finding those pairs of vertices (where there is not already an edge) which are most likely to form an edge in the next time step. 
Any similarity score on pairs of vertices can naturally be considered a link prediction algorithm:  two vertices with a high similarity score are assumed to be more likely to have a link in the future.  Liben-Nowell and Kleinberg~\cite{Liben-Nowell:2003} give a detailed list of such similarity scores. We will also need similarity scores to create ground-truth synthetic data, as detailed in Section~\ref{sec:models}. Let $\text{score}(x,y)$ denote the similarity score between two vertices and $\Gamma(x)$ the neighborhood of $x$.  With this notation, we use the following four scores, as given in~\cite{Liben-Nowell:2003}:

\begin{enumerate}
\item{Adamic-Adar:} $\text{score}(x,y) := \sum_{z\in \Gamma(x)\cap\Gamma(y)} \frac{1}{\log |\Gamma(z)|}$
\item{$\text{Katz}_\beta$:} $\text{score}(x,y) := \sum_{\ell=1}^\infty \beta^\ell \cdot |\{\text{paths of exactly length }\ell\text{ from }x\text{ to }y\}|$
\item{Graph distance:} $\text{score}(x,y) := -d(x,y)$, where $d(x,y)$ is the distance between $x$ and $y$.
\item{Rooted $\text{PageRank}_\alpha$:}  Consider a random walk on the graph that resets to vertex $u$ with probability $\alpha$ and moves to a random adjacent vertex with probability $1-\alpha$.  Then $\text{score}(x,y)$ is the stationary probability of $y$ when resetting to $x$ plus the stationary probability of $x$ when resetting to $y$.
\end{enumerate}

To test the quality of a window size $w$, we analyze the performance of a link prediction algorithm on the graph sequence aggregated at $w$.  The link prediction algorithm gets as input the graph at time $t$ and predicts new edges in the graph at time $t+1$.  We predict the edges with the top $k$ scores, where $k$ is the number of edges that actually are created in the next time step in the sequence.  As is common, we view this as a binary classification task, where pairs of vertices are either in the category of new edges or not.  When viewed in this way, it is natural to score the algorithm as the correlation between the algorithm's predicted edges and those edges that actually appear, which is called the Matthews correlation coefficient\footnote{The MCC measure is used, rather than accuracy or precision, because MCC skirts the issue of bias that accuracy and precision have: the number of edges appearing is often a very small fraction of the total number of possible edges.  We use MCC over other measures also resistant to unequally sized categories, such as AUC, because MCC is computationally very fast.  In addition, it seems to emphasize differences in scores better than other measures for our link prediction task.  Regardless, other measures give very similar scores as AUC - different measures seem to preserve the order of the qualities of the window sizes.}(MCC)~\cite{baldi2000assessing}.  Given a pair of consecutive graphs $G_i$ and $G_{i+1}$ from a graph sequence aggregated at a window size $w$, the Matthews correlation is the $[-1,1]$-valued Pearson correlation coefficient for classification defined as a normed $\chi^2$ statistic between the predicted edges to appear in $G_{i+1}$ (using $G_i$ as the input to the link prediction algorithm) and the actual new edges appearing in $G_{i+1}$.  The score assigned to $w$ is the average of these correlations over all pairs of consecutive graphs in the aggregated sequence.




We test all window sizes\footnote{To be more precise, we only test window sizes up to a third of the length of the input.  We assume that if the actual window size is any bigger, then there is no temporal information that we can utilize and the underlying network is really a static network.}.  This algorithm is summarized in Algorithm~\ref{alg:assign_scores}.

\begin{algorithm}
	\caption{Assign scores to each window size}
	\label{alg:assign_scores}
\begin{algorithmic}
	\STATE $\mathcal{G} = G_1,\ldots, G_n$
	\FOR{$w=1$ to $\lfloor n/3 \rfloor$}
		\STATE Let $\mathcal{G}'$ be $\mathcal{G}$ aggregated at size $w$, so $\mathcal{G}' = G'_1, \ldots, G'_{\lfloor n/w \rfloor}$.
		\FOR{$i=0$ to $|\mathcal{G}'|-1$}
			\STATE new links $= E(G'_{i+1})\setminus E(G'_i)$
			\STATE predicted links $= LP(G'_i, G'_{i+1})$
			\STATE $\text{score}_i = MCC(\text{new links},\text{predicted links})$
		\ENDFOR
		\STATE $\text{score}_w = \frac{\sum_{i=0}^{|\mathcal{G}'|-1} score_i}{|\mathcal{G}'|}$ 
	\ENDFOR
	\RETURN all pairs $w, \text{score}_w$
\end{algorithmic}
\end{algorithm}

Since this is a computationally-expensive task, for sufficiently long sequences, a random ten percent of the consecutive pairs of aggregated graphs are tested instead of all of the pairs. Based on our empirical analysis, considering only a subsample of the windows does not  significantly affect the scores.

Depending on the application, it may be desirable to use more than one different window size - more than one time scale may be interesting or perform highly.  In addition, for a given time scale, there may be a range of window sizes centering around that time scale that is of interest.  Moreover, we leave for future work determining how significant differences in quality scores are, given that we make no assumptions on how many different time scales are of interest for the application.  In this light we do not return just the top-quality window size, as seen in Algorithm~\ref{alg:assign_scores}.

\section{Generative Models for Graph Sequences}\label{sec:models}

To create synthetic data, we will use two generative models, one for the ground-truth sequence representing the noiseless dynamic network, and the other, a noise model that takes as input a ground-truth sequence (which we sometimes refer to as the underlying sequence) and outputs a noisy oversampled sequence.

\subsection{Generative Model for the Ground-truth Graph Sequence}
The ground-truth graph sequence can be formed from any existing model of a dynamic network (such as the latent-space model or the activity-driven model mentioned in the introduction), but to test the performance of a link-prediction-based approach, we instead use a novel and simple generative process that allows us to test our approach, which has the advantage that the quality of the link prediction algorithm is a parameter of the generative process.  This generative process starts with an initial graph $G$ and adds a fixed but parameterized number of edges $\delta$ for every subsequent graph. The edges added are the non-edges with the top scores as rated by a given similarity score. 
This model can easily be extended as needed, for example by deleting the edges with the lowest similarity score every time step as well, or still further to a probabilistic edge creation and deletion process.
 
For the initial graph $G$, in this paper we consider both the Erd\H{o}s-R\'enyi model $G(n,p)$~\cite{erdos1959random} and the preferential attachment model $BA(n,m)$~\cite{barabasi1999emergence}. For the similarity scores, we use Adamic-Adar and Katz. The use of different similarity scores allows us to test our approach both when the quality of the link prediction algorithm does well and when it does not do well. For example, if we use Adamic-Adar to both create the graph sequence and to do link prediction (assuming no noise) the link prediction algorithm will perform perfectly.  However if the sequence is instead made with the Katz similarity score, the link prediction algorithm will not perform as well.  Since we can't guarantee the quality of the link prediction algorithm on non-synthetic data, this model allows us to see how link prediction performance affects our approach.

\subsection{Generative Model for Oversampling}
We now need to model the sampling process by which non-synthetic data would be gathered.  Our primary approach to modeling noise, specifically oversampling, is to assume that for a given time step, the edges that occur are measured to be near that time step, but not necessarily at that time step.  Furthermore, we assume the distribution of these edges in time is Gaussian.  Given an input graph sequence of length $t$ and parameters $\mu\in\mathbb{N}, \sigma^2 \in \mathbb{R}_{\ge 0}$, this model outputs a graph sequence of length approximately $\mu\cdot t$ that represents the sequence being oversampled at a constant rate $\mu$, with $\sigma^2$ controlling how concentrated the edges occur around the ``true" times. Specifically, given these two parameters $\mu$ and $\sigma$, for an edge that occurs in the ground-truth graph at time step $i$, the edge will occur in the new noisy graph sequence at time step $j\sim\mathcal{N}(\mu i,\sigma^2)$, where $j$ is rounded to the nearest integer. 
 
If $\sigma$ is sufficiently small, then there is likely to be intermediate graphs where there are no edges.  If windows start and end within these gaps, we can recover fully from this noisy process.  However, as $\sigma$ gets larger, it becomes more and more difficult, as more and more edges from distinct graphs in the original network start getting added to the same graphs in the new noisy network.  In the limit, all temporal information is destroyed.  Figure~\ref{fig:AAgaussian1-numedges} gives two examples of the number of edges in each time step for two different settings of parameters.   
It's worth noting that the oversampling noise model presented here can be extended so that the oversampling rate is non-constant, the distribution used is non-Gaussian, etc.

\section{Results for Synthetic Data}\label{sec:gaussiannoise}
In this section, we analyze a variety of oversampled, dynamic networks generated by models detailed in Section~\ref{sec:models} and aggregated at different window sizes. We investigate the effect of window size on the performance of link prediction.  We show that link prediction performs better on sequences aggregated at window sizes close to ground-truth than on other window sizes.  Furthermore, this holds when the ground-truth sequence uses a different similarity score than the one used to perform link prediction.  We also show that  larger values of $\mu$, while fixing $\sigma$ (increasing the separation between means), smaller values of $\sigma$, while fixing $\mu$ (higher degree of concentration around each mean), larger values of $\delta$ (more edges appearing each time step), and higher quality of link prediction all make the task of recovering the ground-truth window size easier.  Yet even at higher levels of noise we show that this approach recovers reasonable results and gives some evidence that it outperforms the simpler approaches that rely solely on extracted time series information, such as the number of edges per time step.

\begin{figure}[!htbp]
\begin{center}
\includegraphics[scale = 0.28]{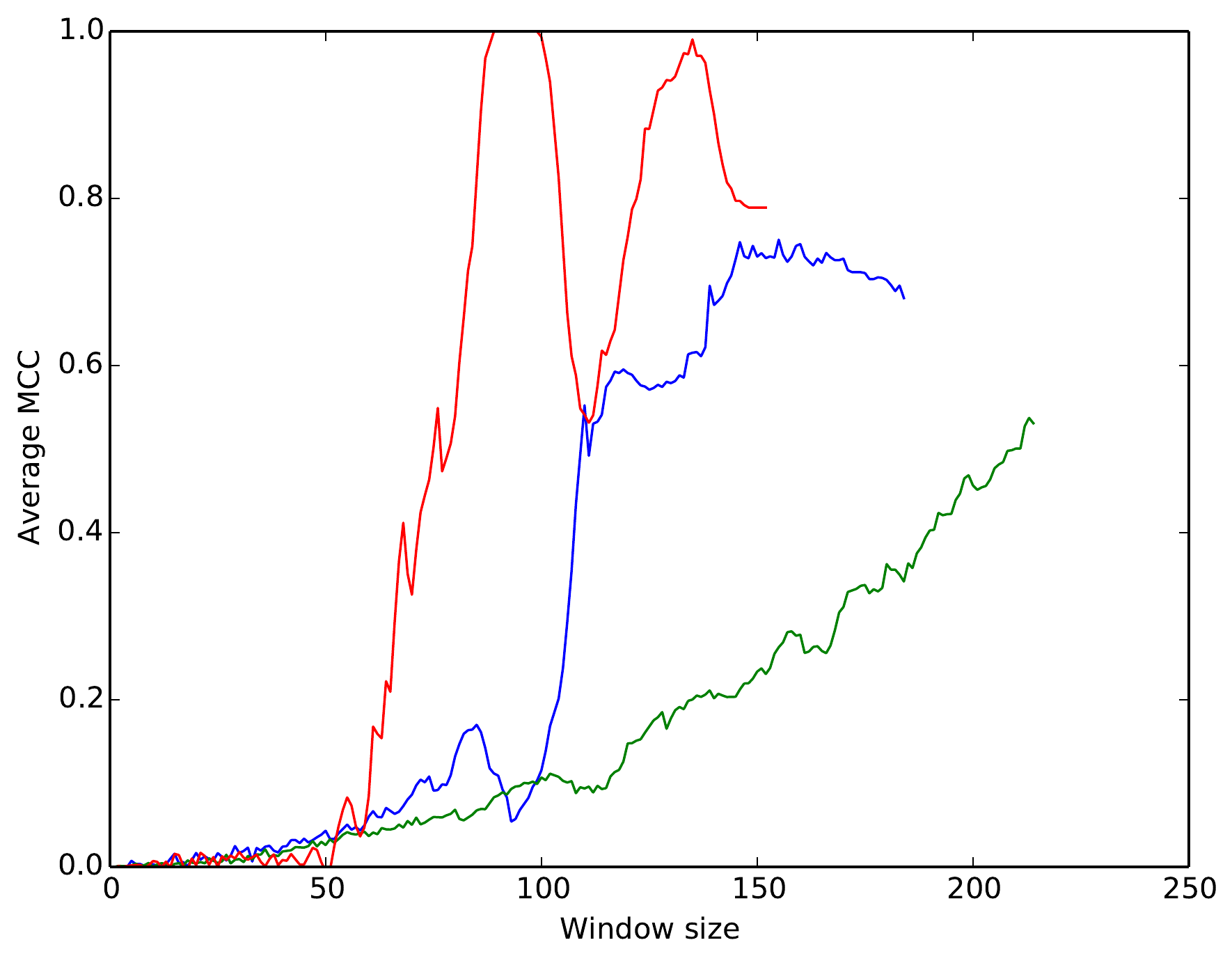}
\end{center}
\caption{MCC scores as a function of window size for three different noisy sequences.  Parameters used were $\mu=100$, $\delta=50$, and from shortest to longest, $\sigma=8,20,40$.  The underlying graph sequence is generated by starting graph $G(n,p)$ and Adamic-Adar as the similarity score.}
\label{fig:AAgaussian1}
\end{figure}

For the sake of brevity, given the number of parameters, including $\mu$, $\sigma$, $\delta$, the number of link prediction algorithms both for creating the sequence and for finding the quality scores, we do not show results for all possible combinations, but instead show a representative sample.



In the remainder of this section, we fix the number of vertices $n$ to be $250$ and fix the edge probability $p$ for the Erd\H{o}s-R\'enyi model at $0.05$. Figure~\ref{fig:AAgaussian1} shows the drastic improvement of link prediction performance at larger windows of aggregation. For the sequence generated using $\sigma=8$, as shown in this figure, performance at window size of 1 (no aggregation) is essentially random (average $\text{MCC}\approx -10^{-4}$), but when aggregated at a window size of 95, performance is perfect (average $\text{MCC}= 1.0$).  Note that the best performing window size is very close to the mean separation $\mu=100$. When the standard deviation is comparatively higher, as when $\sigma=40$, Figure~\ref{fig:AAgaussian1} shows that windowing has a comparatively smaller effect.  Here the link prediction is not able to separate consecutive graphs and therefore recommends to aggregate all graphs in the sequence together.
 
 \begin{figure}[!htbp]
 \begin{center}
 \begin{subfigure}[b]{0.45\textwidth}
                \includegraphics[width=\textwidth]{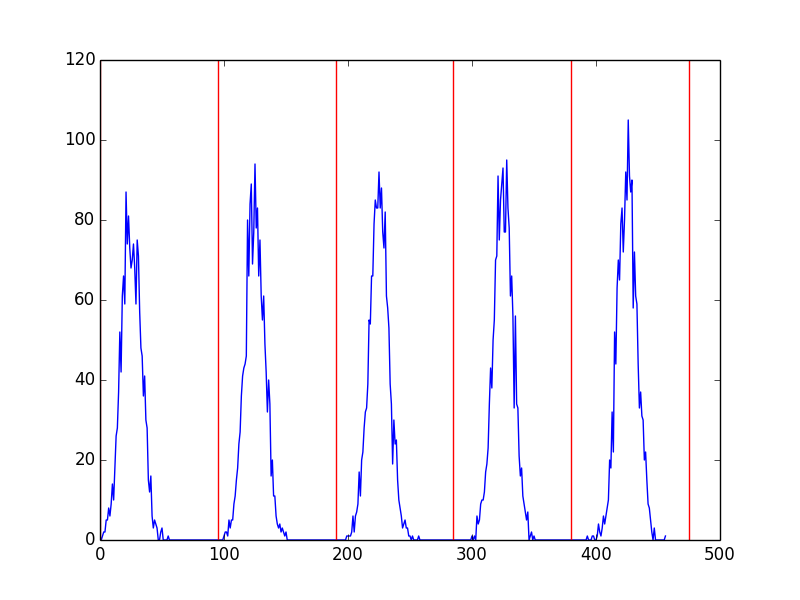}
                \caption{$\mu=100$, $\delta = 50$, $\sigma=8$}
                \label{fig:seqs-19-50-numedges}
\end{subfigure}
\begin{subfigure}[b]{0.45\textwidth}
                \includegraphics[width=\textwidth]{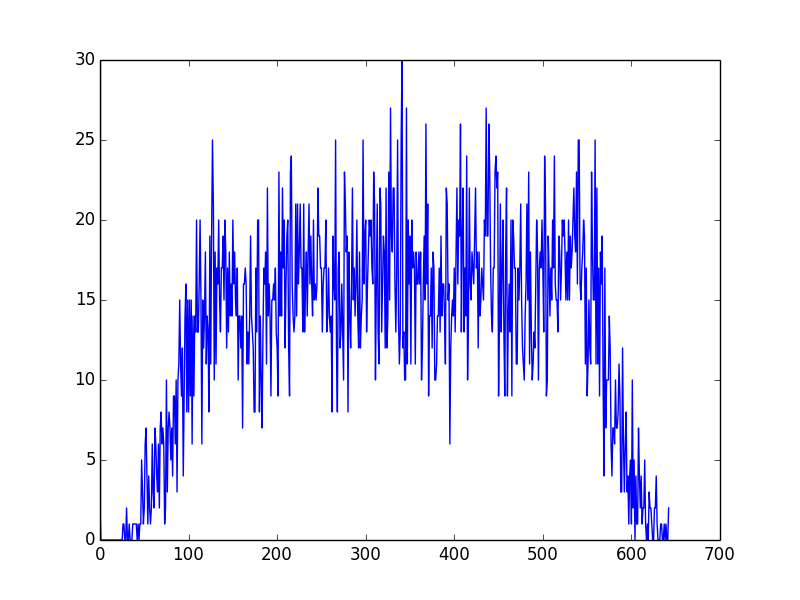}
                \caption{$\mu=100$, $\delta = 50$, $\sigma=40$}
                \label{fig:seqs-19-56-numedges}
\end{subfigure}
\caption{Number of edges in two noisy sequences over time with the given parameters.  The underlying sequences were created with $G(n,p)$ as the starting graph and Adamic-Adar as the similarity score. The vertical lines indicate the borders of windows when the window size is 95.}
\label{fig:AAgaussian1-numedges}
\end{center}
 \end{figure}
 
Figure~\ref{fig:AAgaussian1-numedges}  gives further evidence for why this is the case.  When $\sigma=40$, mixing (of edges) between graphs is much higher than in the case when $\sigma=8$.  In this latter case, there is in fact no mixing at all, which is why there are many window sizes where link prediction performs perfectly.  Since link prediction here is so good - it will be perfect in the absence of noise - it represents an easier case. However, this case still demonstrates a convenient benefit to our approach: we can still perform well even when the target windowing is not uniform over the length of the input noisy sequence. That is, the noisy sequence may have shifted windows and as a result, one window could be smaller than the others. In our case, window size 95 performs better than a window size 100, indicating we can still find a \emph{uniform} window size that performs well, even if it is not the same window size as the mean separation window.  

\begin{figure}[!htbp]
	\begin{center}
		\begin{subfigure}[b]{0.45\textwidth}
			\includegraphics[width=\textwidth]{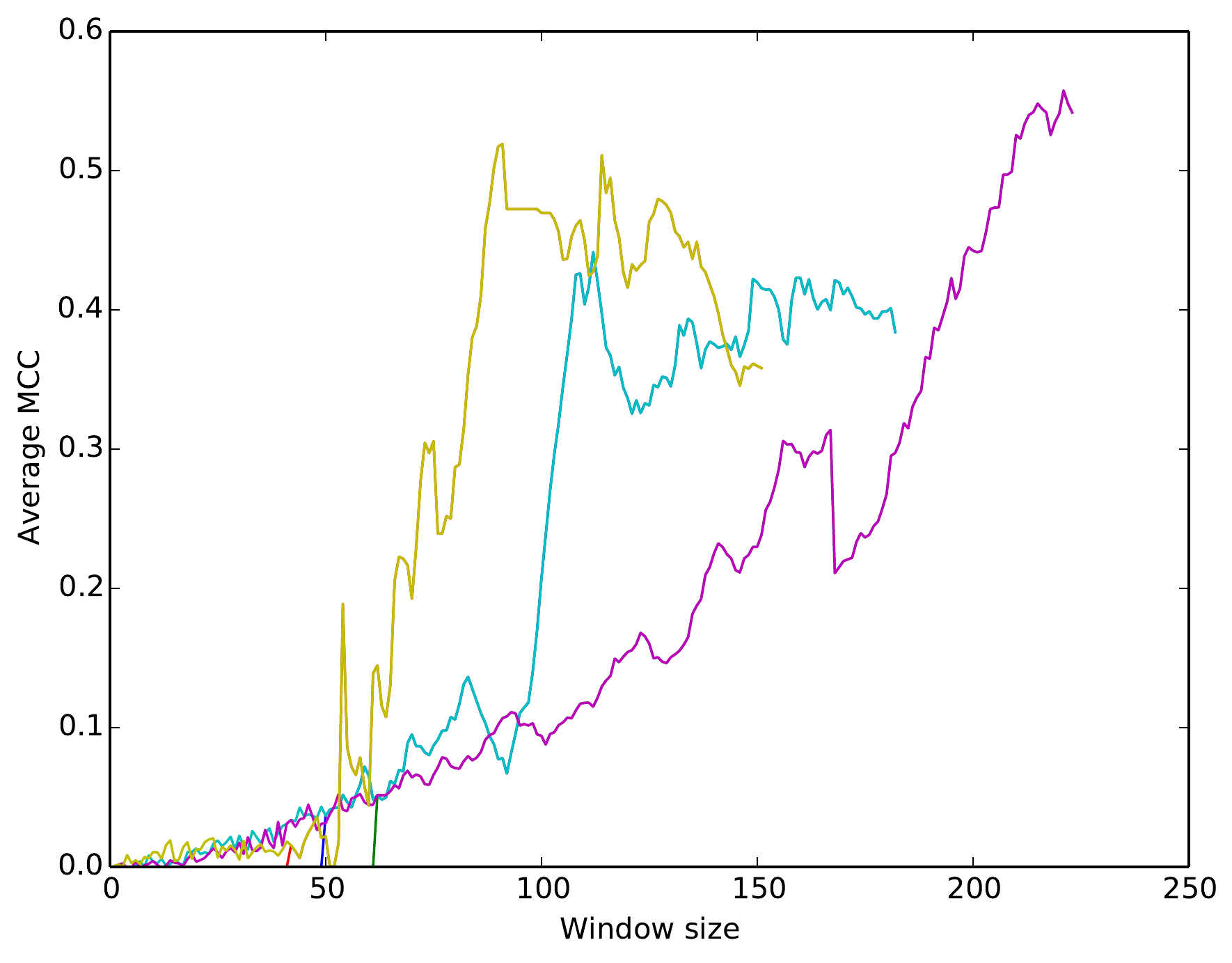}
			\caption{$\mu=100$, $\delta=50$, and from shortest to longest, $\sigma=8$, $20$, and $40$}
			\label{fig:Katzgaussian1}
		\end{subfigure}
		\begin{subfigure}[b]{0.45\textwidth}
			\includegraphics[width=\textwidth]{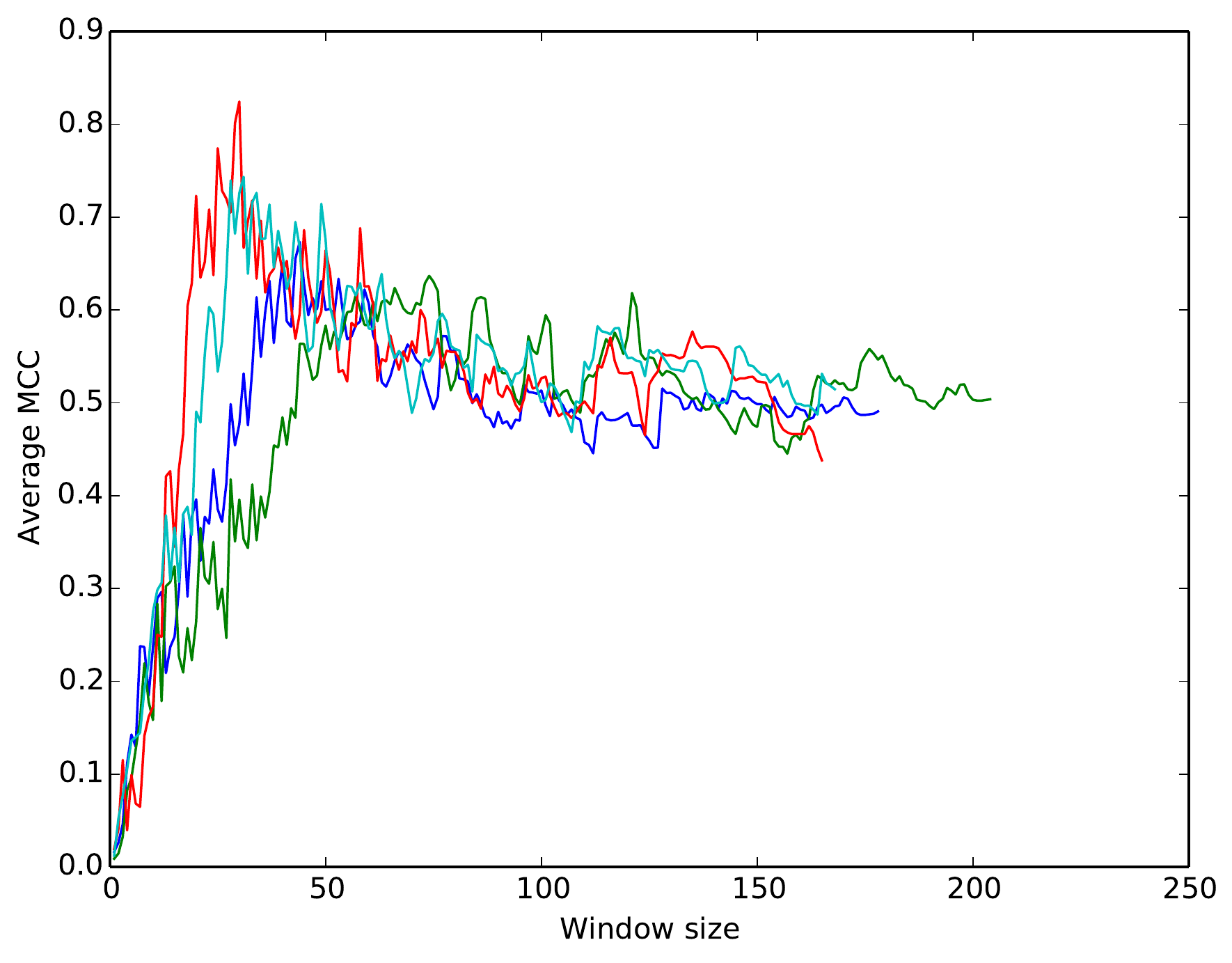}
			\caption{$\mu=20$, $\delta=50$,  and $\sigma=2,4,8,20$ \hspace{2cm} }
			\label{fig:Katzgaussian2}
		\end{subfigure}
		
		\begin{subfigure}[b]{0.45\textwidth}
			\includegraphics[width=\textwidth]{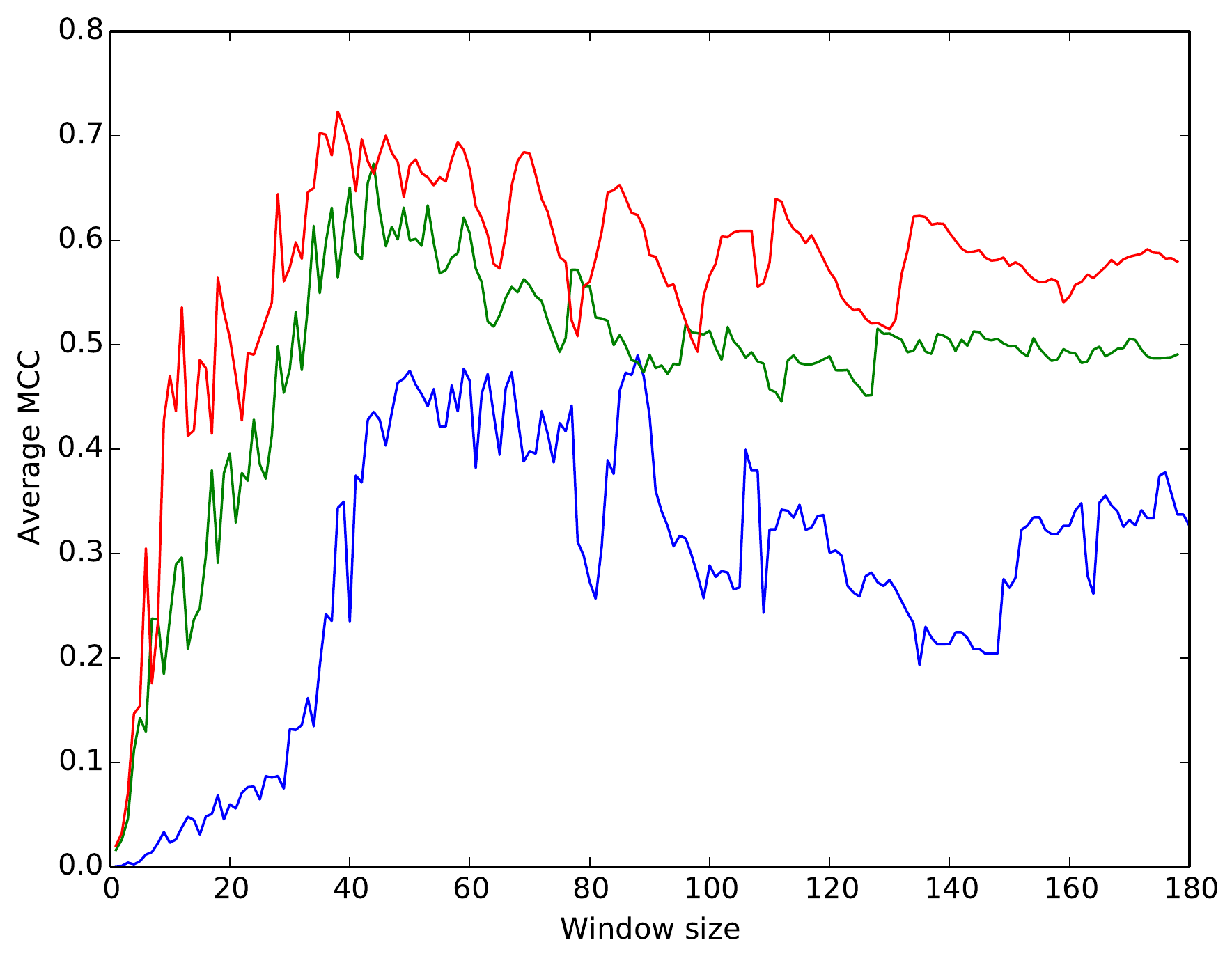}
			\caption{$\mu=20$, $\sigma = 8$, and, from top to bottom, $\delta=100$, $50$, and $5$}
			\label{fig:Katzgaussian3}
		\end{subfigure}
		\begin{subfigure}[b]{0.45\textwidth}
			\includegraphics[width=\textwidth]{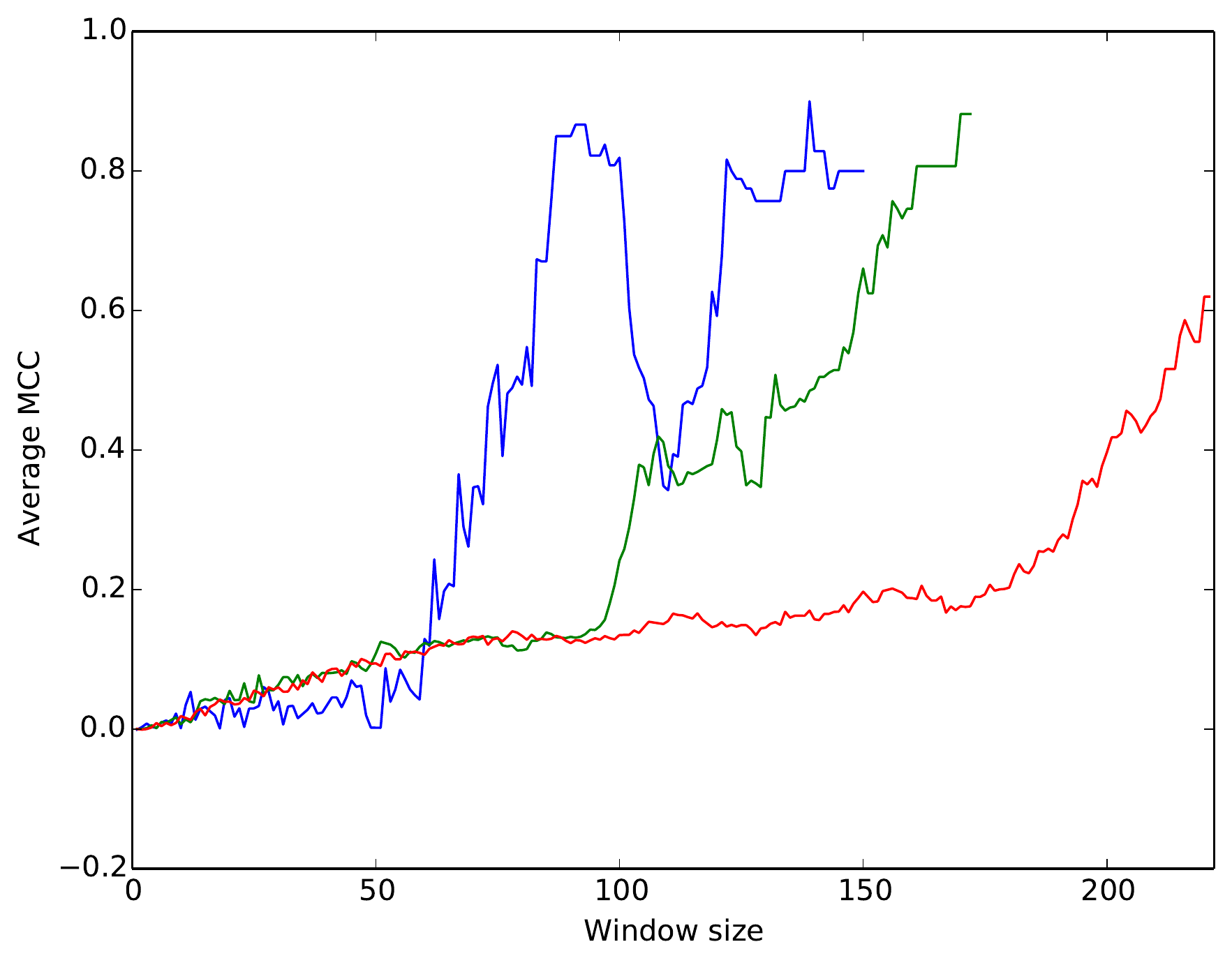}
			\caption{$\mu=100$, $\delta = 5$, and, from shortest to longest, $\sigma=8$, $20$, and $40$}
			\label{fig:Katzgaussian4}
		\end{subfigure}
	\end{center}
	\caption{MCC scores as a function of window size for different noisy sequences where the Katz similarity score was used to generate the underlying sequence.  In Figure~\ref{fig:Katzgaussian4}, the starting graph is the preferential attachment graph $BA(n,5)$ instead of $G(n,p)$.}
\end{figure}

Figure~\ref{fig:Katzgaussian1} illustrates a similar behavior when the ground truth sequence was instead created with the Katz similarity score (testing of each window size is still done using the Adamic-Adar score as the link prediction algorithm).  While overall quality scores are lower, as the link prediction algorithm itself is worse, the same pattern holds: near a window size of 100, the link prediction algorithm does significantly better. 

Our approach is more resilient to a higher value of $\sigma$, as seen in Figure~\ref{fig:Katzgaussian2}. However, even when edge mixing is very high ($\sigma=20$ and $\mu=20$), 
link prediction is still helpful in identifying a good window size (near $\mu = 20$).  This gives significant evidence that our approach outperforms simpler approaches that rely solely on extracted time series information such as the number of edges per time step.  Finally, link prediction does better at recovering a good window size when $\delta$ is higher - it retains its resiliency even for very sparse cases, as seen in Figure~\ref{fig:Katzgaussian3}. 


As mentioned above, our generative model easily extends to more general models.  Again, for the sake of brevity, we will not discuss the performance of our approach on every possible variation of our model, but we do want to note that our results do extend.  For example, Figure~\ref{fig:Katzgaussian4} shows our results when the following two changes have been made:  The initial graph is instead of $G(n,p)$ an instance of the preferential attachment model.  In addition, instead of just adding edges, edges are deleted as well.  Namely, $\delta$ existing edges that have the lowest Katz score are deleted.  These changes show very little impact on our results, as seen by comparing Figures~\ref{fig:AAgaussian1} and~\ref{fig:Katzgaussian1}.

\section{Results for Real-world Data}\label{sec:real data}

The validation process on real-world data is difficult, in general, for inference tasks on networks, but especially for the problem of temporal scale identification.  A significant barrier is the lack of ground truth and/or formal notions of what should be considered a good  temporal scale. Real-world dynamic networks often exhibit multiple critical temporal scales corresponding to the evolution of different important features (e.g. communities) and processes (e.g random walks)~\cite{sulo2010meaningful,ribeiro2013quantifying}, but the relationship between these important features and processes and their corresponding temporal scales is not well understood. 

Our expectation is that there can be multiple peaks in the quality of a link prediction algorithm as window size increases, each corresponding to different important temporal scales.
We validate our results by making sure that different link prediction algorithms behave similarly as a function of window size, despite following different mechanisms for scoring future edges.  When two different link prediction algorithms show peaks 
at the same time scale despite not necessarily predicting the same edges, this gives evidence that the peaks are inherit to the sequence and not a function of the particular algorithm.
We now show how our approach performs on real-world data sets, namely the Haggle Infocom network~\cite{cambridge-haggle-2006-01-31}  and the MIT Reality Mining network~\cite{eagle2006reality}. 

\subsection{Haggle Infocom}
\begin{figure}[htbp]
\begin{center}
\begin{subfigure}[b]{0.4\textwidth}
                \includegraphics[width=\textwidth]{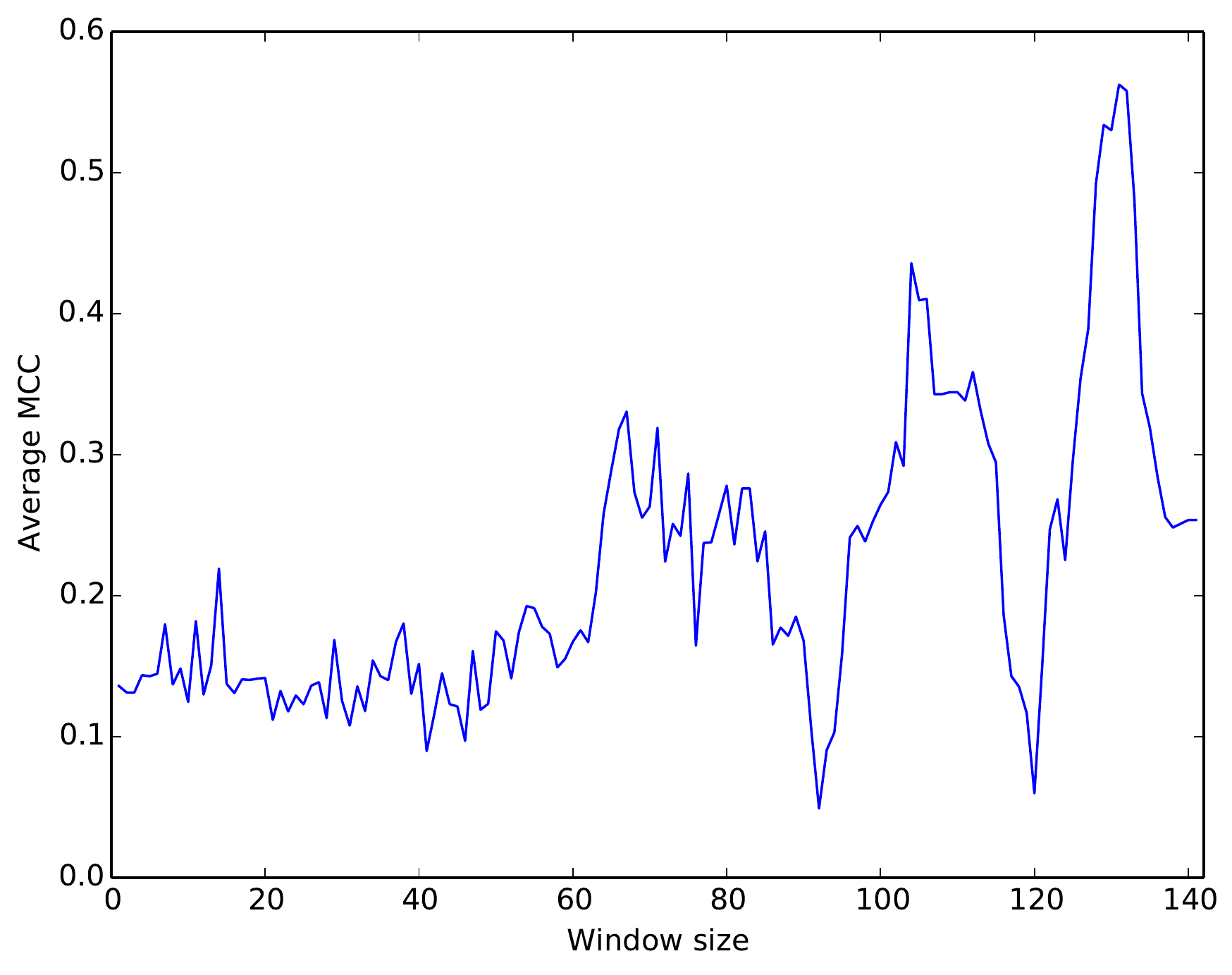}
                \caption{Adamic-Adar}
                \label{fig:haggle_aa}
\end{subfigure}
\begin{subfigure}[b]{0.4\textwidth}
                \includegraphics[width=\textwidth]{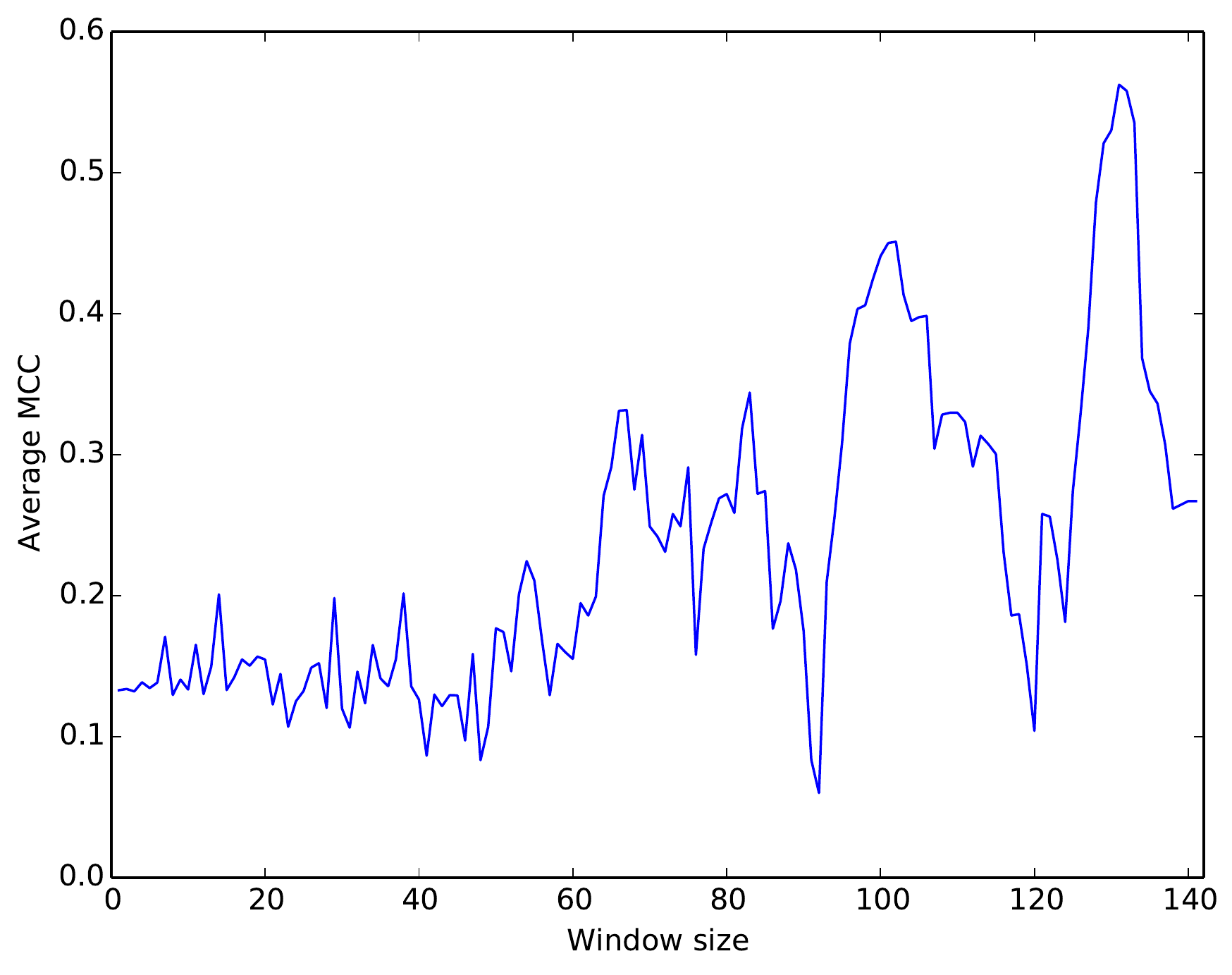}
                \caption{$\text{Katz}_\beta$ ($\beta=0.005$)}
                \label{fig:haggle_katz}
\end{subfigure}
\begin{subfigure}[b]{0.4\textwidth}
                \includegraphics[width=\textwidth]{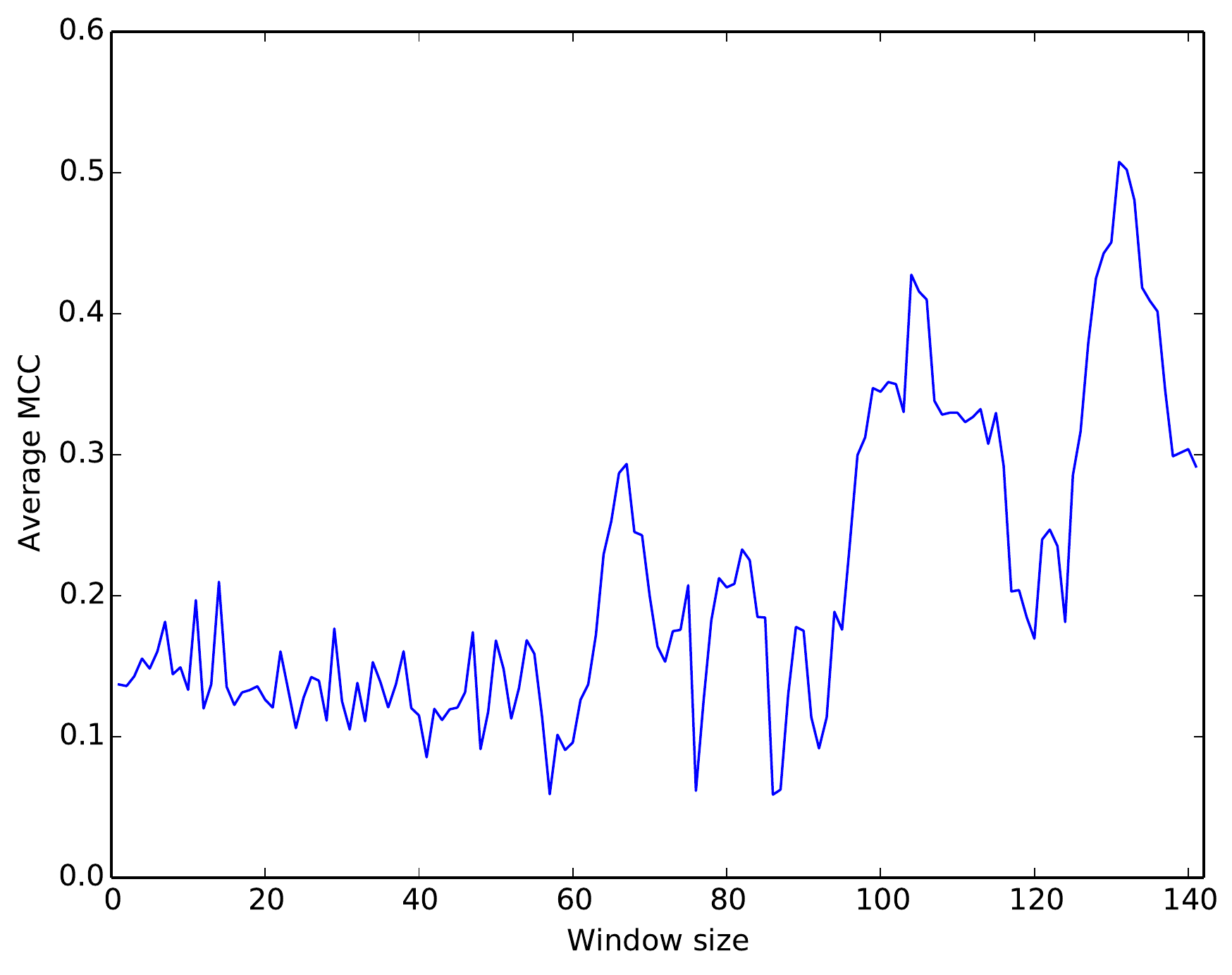}
                \caption{Rooted $\text{PageRank}_\alpha$ ($\alpha = 0.15$) }
                \label{fig:haggle_rpr}
\end{subfigure}
\begin{subfigure}[b]{0.4\textwidth}
                \includegraphics[width=\textwidth]{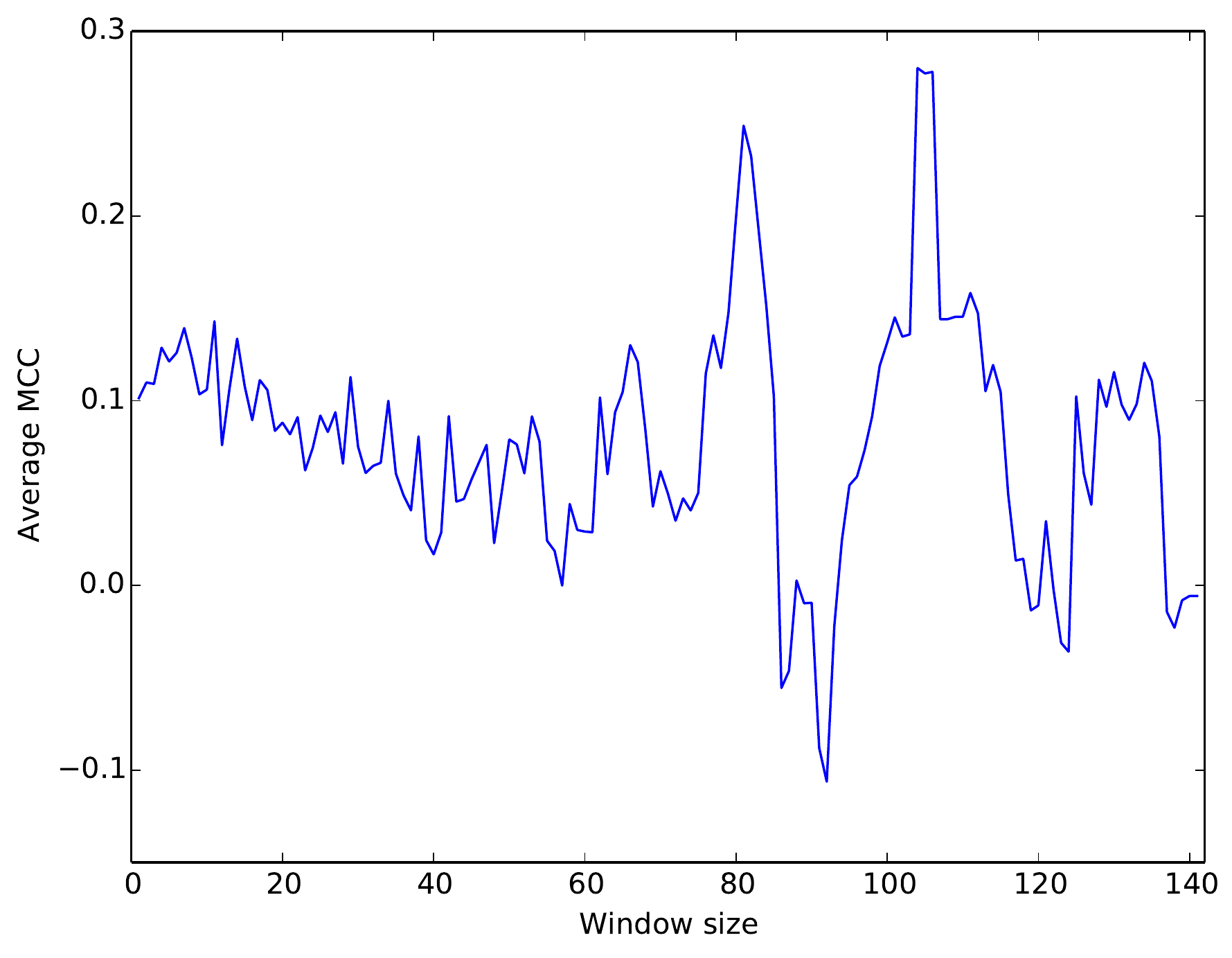}
                \caption{Graph distance}
                \label{fig:haggle_gd}
\end{subfigure}

\caption{MCC scores for four link prediction algorithms as a function of window size for the Haggle sequence.}
\label{fig:haggle}
\end{center}
\end{figure}

The Haggle Infocom dataset is the result of 41 users equipped with Bluetooth phones at the Infocom 2005 conference, over the course of four days.  There is an undirected edge between two users at time $t$ if they were in proximity at that time. The data we used was initially binned at 10 minute time intervals.  The results are shown in Figure~\ref{fig:haggle}. The four link prediction algorithms behave consistently with clear peaks at approximately $w=75$ ($\approx 12.5$ hours), $w=110$ ($\approx 18$ hours), and $w=130$ ($\approx 22$ hours). The interactions present in the Haggle network have a periodic nature imposed by the regular conference structure and our algorithm seems to identify such periodicities, in particular the half and one-day periodicities. 

\subsection{MIT Reality Mining}

The MIT Reality Mining dataset  consists of 90 grad students and professors' data from their cell phones in the 2004-2005 academic year.  Timestamps were kept for three types of data:  Bluetooth proximity, cell tower proximity and phone call communications. This naturally yields three different dynamic networks that we extracted from the raw data. The first network has an undirected edge at time $t$ whenever Bluetooth recorded two cell phones as close at time $t$, the second has an undirected edge between two participants at time $t$ if they were recorded near the same cell tower, and finally the third has an undirected edge between two participants at time $t$ whenever one participant called the other.  
We will refer to these as the Bluetooth sequence, the cell tower sequence, and the call sequence respectively.  Each network is initially windowed at a size of one day.

\begin{figure}[!htbp]
\begin{center}
\begin{subfigure}[b]{0.4\textwidth}
                \includegraphics[width=\textwidth]{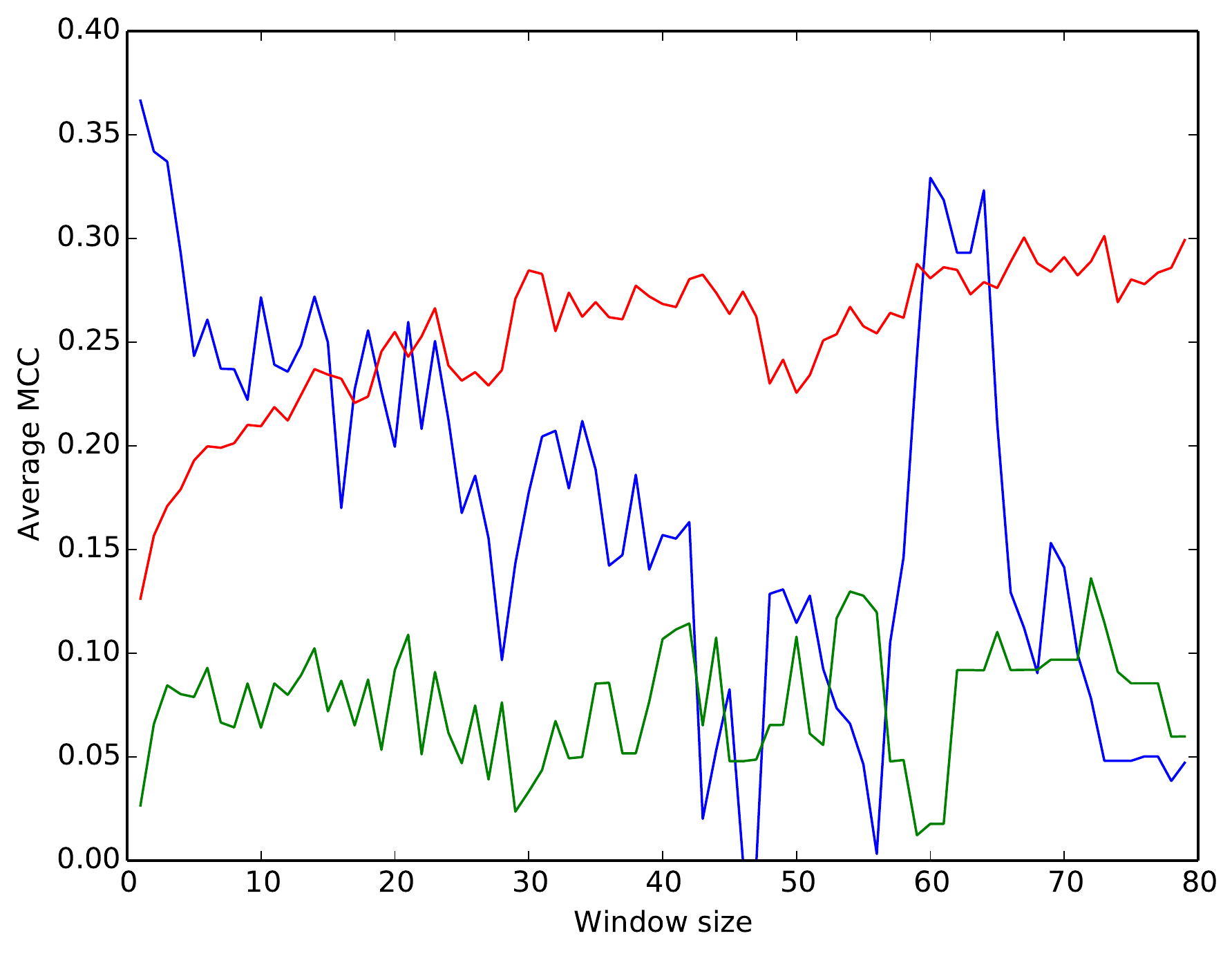}
                \caption{Adamic-Adar}
                \label{fig:reality_mining_aa}
\end{subfigure}
\begin{subfigure}[b]{0.4\textwidth}
                \includegraphics[width=\textwidth]{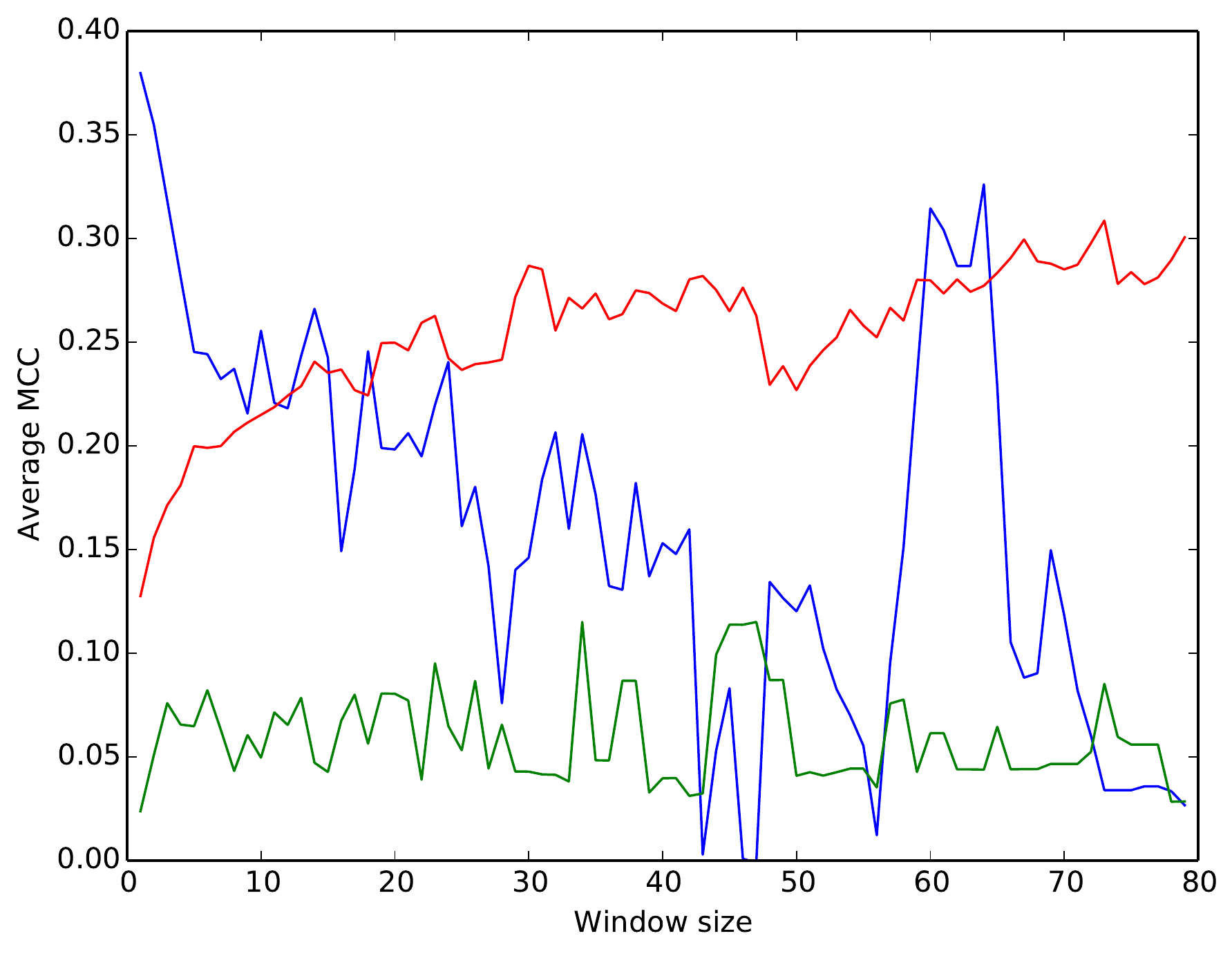}
                \caption{$\text{Katz}_\beta$ ($\beta=0.005$)}
                \label{fig:reality_mining_katz}
\end{subfigure}

\begin{subfigure}[b]{0.4\textwidth}
                \includegraphics[width=\textwidth]{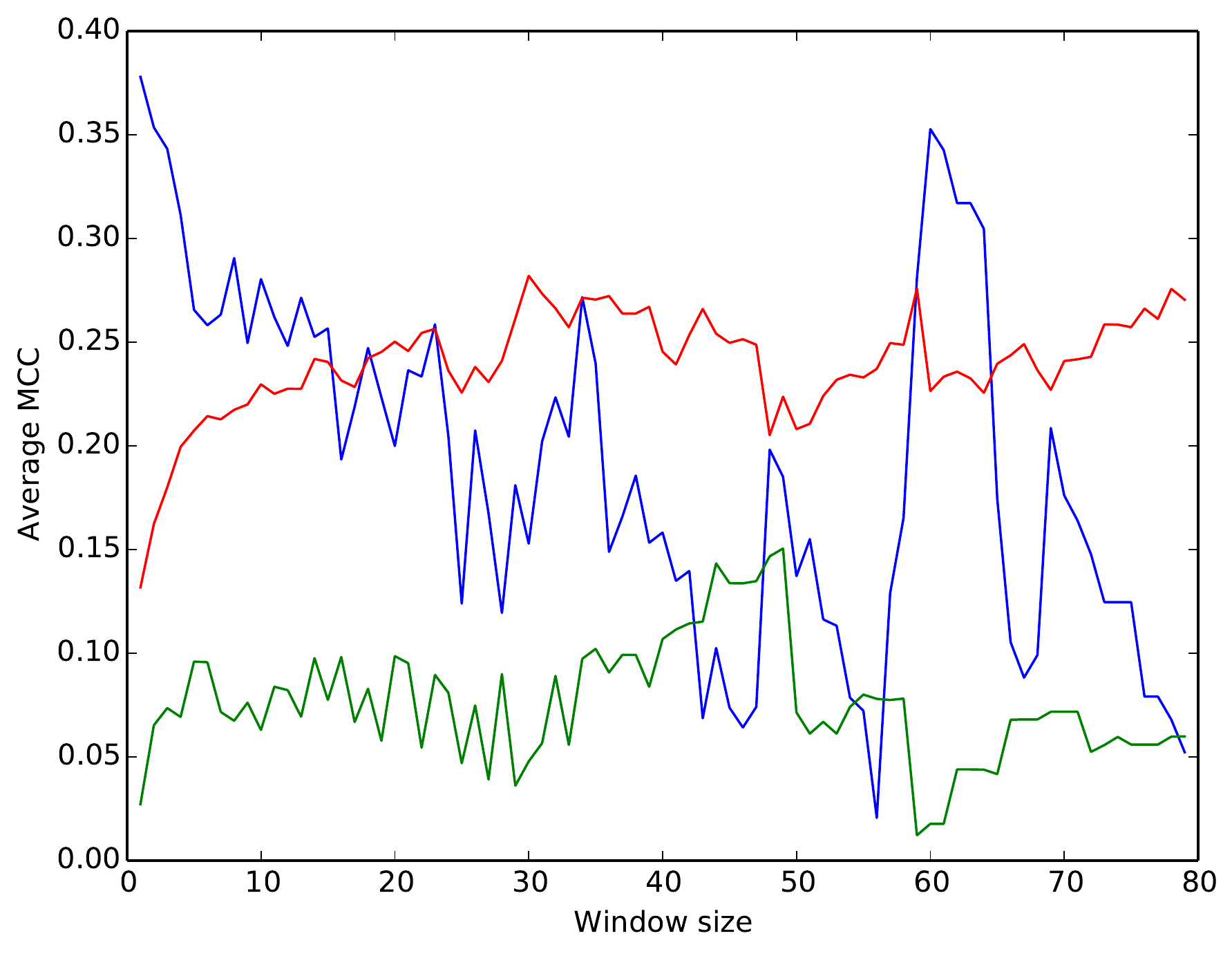}
                \caption{Rooted $\text{PageRank}_\alpha$ ($\alpha = 0.15$)}
                \label{fig:reality_mining_rpr}
\end{subfigure}
\begin{subfigure}[b]{0.4\textwidth}
                \includegraphics[width=\textwidth]{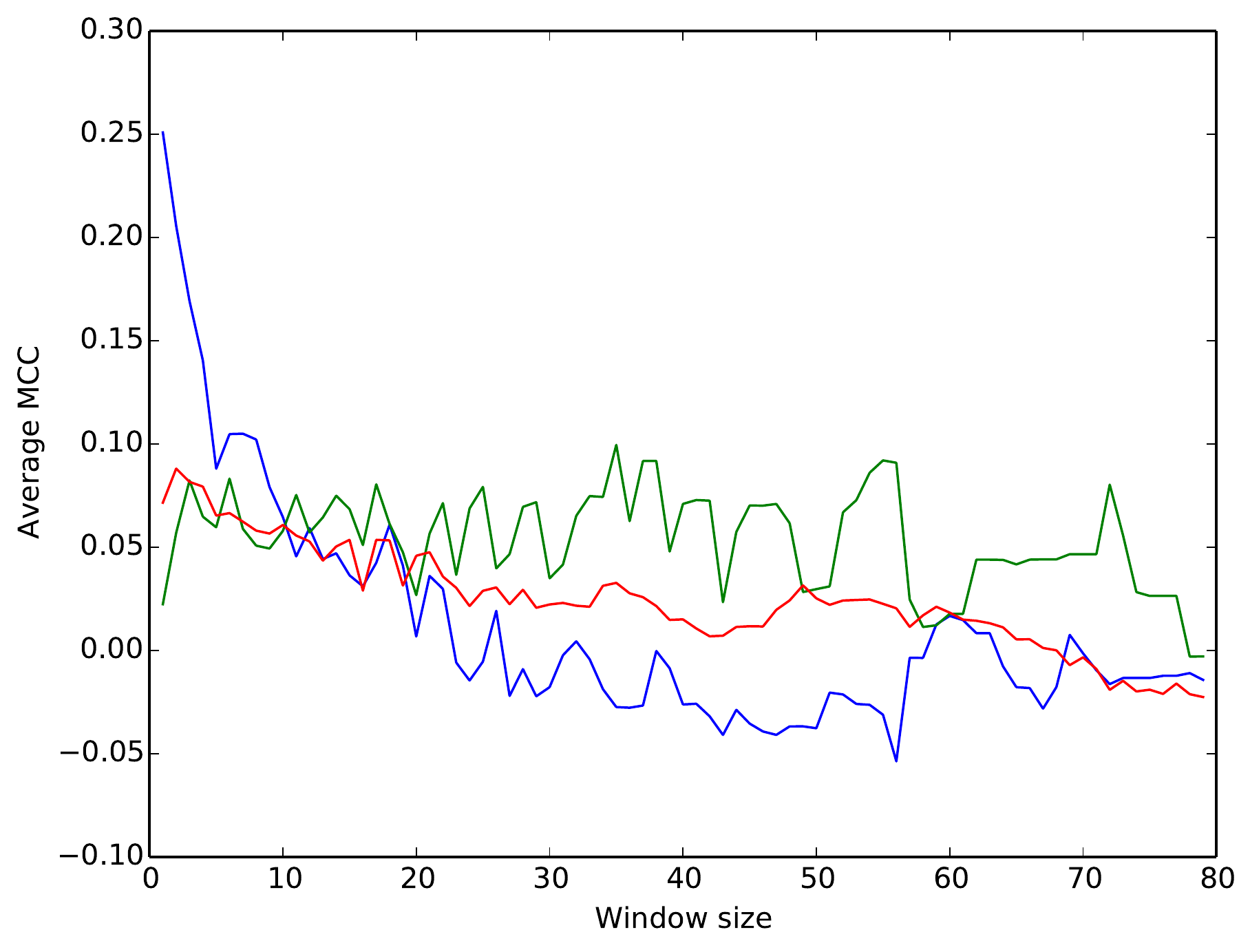}
                \caption{Graph distance}
                \label{fig:reality_mining_gd}
\end{subfigure}

\caption{MCC scores for four link prediction algorithms as a function of window size for the Reality Mining sequence.  At the far left of each plot, from top to bottom is the cell tower sequence, the Bluetooth sequence, and the call sequence respectively.}
\label{fig:reality_mining}
\end{center}
\end{figure}

The results are shown in Figure~\ref{fig:reality_mining}.  Of interest is that the three different networks perform - in terms of the quality of the link prediction - consistently differently from each other, implying that they really are different types of networks. The Bluetooth sequence shows clear signs of oversampling; a window size of $w=1$ has smaller quality than larger window sizes. The performance of link prediction on this sequence stabilizes after $w=14$ (roughly 2 weeks), while in the case of the cell tower sequence, the performance drops and then prominently improves at $w=65$ (roughly 2 months). Considering the weekly, monthly, and semester structure of academic activities, these windows of aggregations appear reasonable in capturing the underlying dynamics of the Reality Mining networks. 

\begin{figure}[!htbp]
\begin{center}
\begin{subfigure}[b]{.47\textwidth}
	\includegraphics[width=\textwidth]{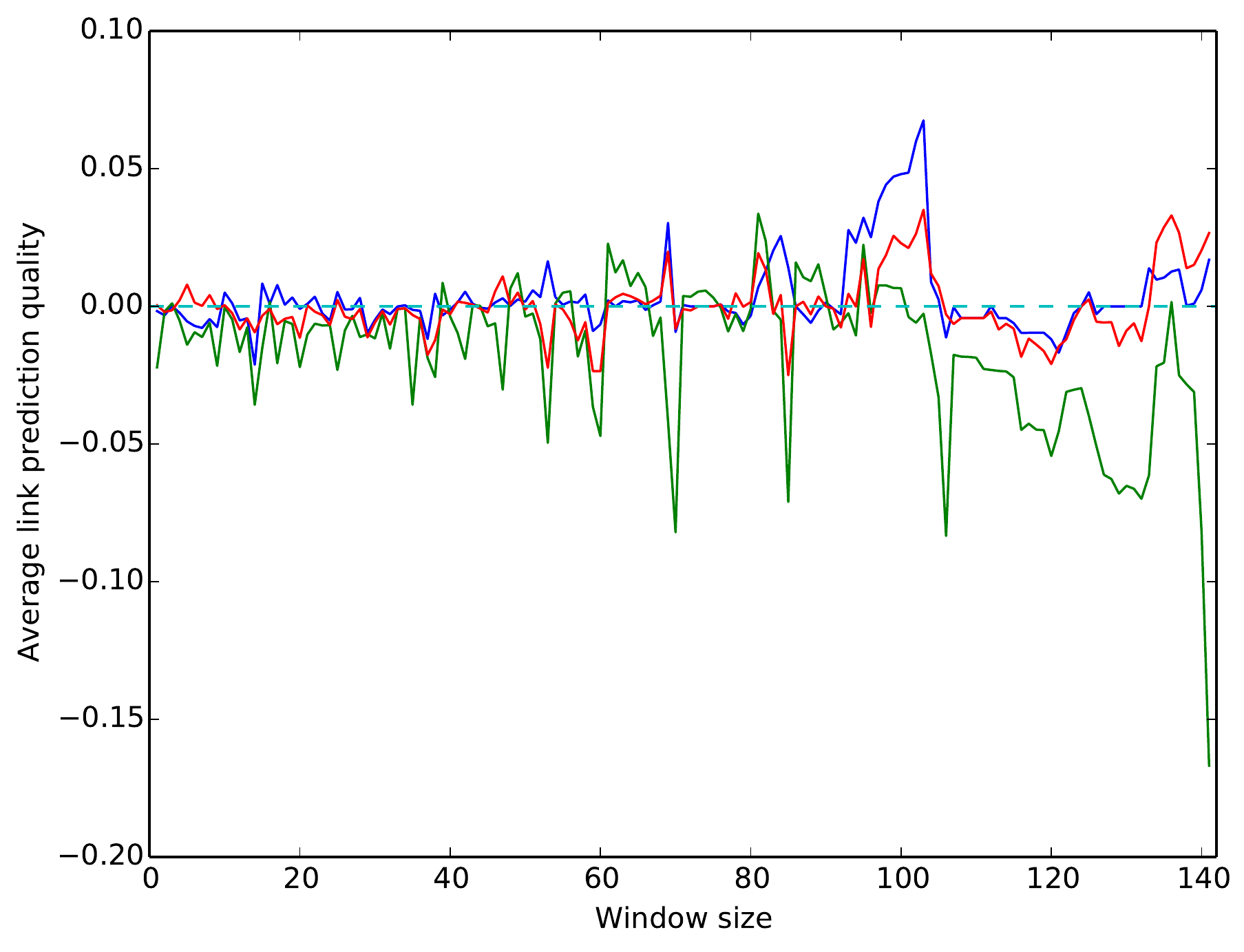}
	\caption{Haggle sequence}
	\label{fig:haggle_resemblances}
\end{subfigure}
\begin{subfigure}[b]{.47\textwidth}
	\includegraphics[width=\textwidth]{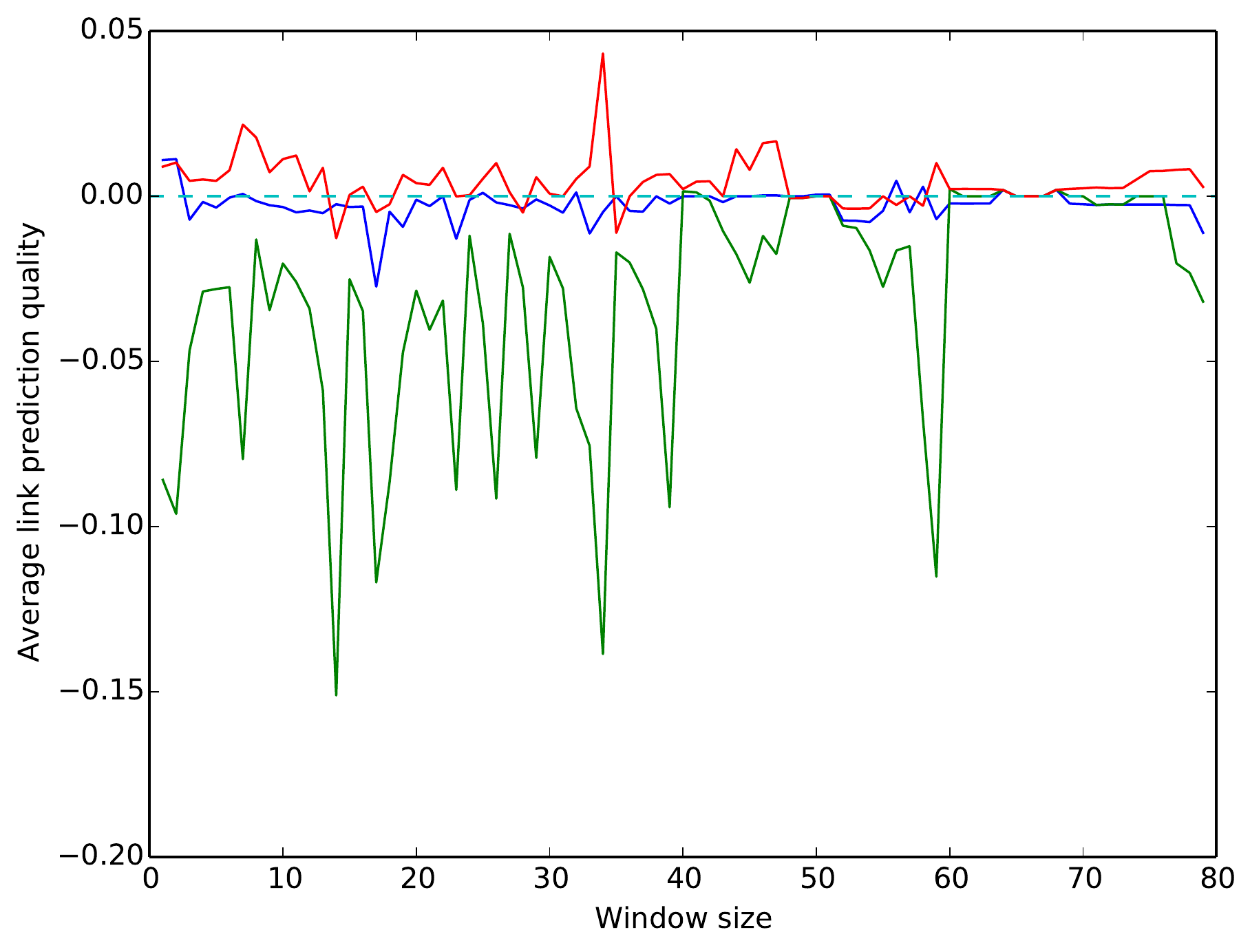}
	\caption{Reality Mining cell tower sequence}
	\label{fig:rm_proximity_resemblances}
\end{subfigure}
\end{center}
\caption{Average quality of the link prediction algorithms Rooted PageRank, Katz, and graph distance, from top to bottom (as seen at furthest to the right) in each plot.  Here the quality of the link prediction is measured as the average resemblance to the actual links that appear, where the resemblance of a set of predicted links to actual set is the size of their intersection divided by the size of their union.  The scores are then normalized so that the resemblance of Adamic-Adar is $0.0$ (the dotted line).}
\label{fig:resemblances}
\end{figure}

These results are given weight by the agreement between three link prediction algorithms, Adamic-Adar, Katz, and Rooted PageRank, as seen in Figure~\ref{fig:reality_mining}. The exception is the graph distance algorithm, whose performance is significantly worse than the other three algorithms, ultimately making it difficult to discern the quality of different window sizes. The graph distance algorithm has been studied in the literature before and is identified as a very low performing link prediction algorithm~\cite{Liben-Nowell:2003}. Figure~\ref{fig:rm_proximity_resemblances} shows the relative performance of the four link prediction algorithms for the Haggle and Reality Mining datasets and re-emphasizes the conclusions in~\cite{Liben-Nowell:2003}. In picking the graph distance algorithm for our problem, we wanted to investigate whether the performance of a bad prediction algorithm can be substantially improved by a better window of aggregation. As the analysis of the Haggle and Reality Mining datasets shows, while some loss in the quality of prediction can be overcome (as in the case of the Haggle sequence), sufficiently bad prediction cannot (as in the case of Reality Mining sequence), as should be expected.


\section{Conclusions}\label{sec:conclusion}
In this work, we treat link prediction as a signal that helps us understand the temporal dynamics of a network. Making the connection between the ability to predict new edges  and the appropriate temporal scale that captures the evolution of the network, we present a novel, task-driven algorithm for de-noising oversampled dynamic network into more robust representations.
We formally define a model of oversampling in dynamic networks that captures general properties of the noise it induces and allows for a more rigorous analysis of our algorithm. 
Across  a variety of synthetic instances of noisy, oversampled dynamic networks, and two real dynamic networks, we show that the performance of a link prediction algorithm can serve as a good quality score for identifying the appropriate window of aggregation. 

Our work opens up several potential avenues for further investigation. As mentioned above, we do not discuss in this paper how to select which windows are outliers in terms of link prediction quality. We would also be interested in extending the framework presented here to non-uniform partitions of the timeline of the network, instead of an inform window of aggregation. 


We only investigate link prediction but other tasks, such as community detection or other inference tasks, may be used to drive the choice of window size. One direction for further work is to investigate how the choice of inference task affects the choice of window size and to what degree do other inference tasks agree with link prediction.

Finally, we leave issues of scalability - both in terms of properties of link prediction as the size of the graphs gets very large and in terms of how to make our algorithmic approach suitably fast for such large graphs - for future work.

\section*{Acknowledgements}
The authors would like to thank Tanya Berger-Wolf for helpful discussions and her generous support.

\bibliographystyle{plain}
\bibliography{time_scale_paper}

\begin{thebibliography}{10}

\bibitem{adafre2005discovering}
Sisay~Fissaha Adafre and Maarten de~Rijke.
\newblock Discovering missing links in {W}ikipedia.
\newblock In {\em Proc. of the 3rd Int. Workshop on Link Discovery}, pages
  90--97. ACM, 2005.

\bibitem{adamic2003friends}
Lada~A. Adamic and Eytan Adar.
\newblock Friends and neighbors on the web.
\newblock {\em Social Networks}, 25(3):211--230, 2003.

\bibitem{airoldi2006mixed}
E.M. Airoldi, David~M. Blei, E.~Xing, and S.~Fienberg.
\newblock Mixed membership stochastic block models for relational data, with
  applications to protein-protein interactions.
\newblock In {\em Proc. of Int. Biometric Society - ENAR Annual Meetings},
  volume~5, 2006.

\bibitem{al2011survey}
Mohammad Al~Hasan and Mohammed~J. Zaki.
\newblock A survey of link prediction in social networks.
\newblock In {\em Social Network Data Analytics}, pages 243--275. Springer,
  2011.

\bibitem{baldi2000assessing}
Pierre Baldi, S{\o}ren Brunak, Yves Chauvin, Claus~A.F. Andersen, and Henrik
  Nielsen.
\newblock Assessing the accuracy of prediction algorithms for classification:
  an overview.
\newblock {\em Bioinformatics}, 16(5):412--424, 2000.

\bibitem{barabasi1999emergence}
Albert-L{\'a}szl{\'o} Barab{\'a}si and R{\'e}ka Albert.
\newblock Emergence of scaling in random networks.
\newblock {\em Science}, 286(5439):509--512, 1999.

\bibitem{clauset2012persistence}
Aaron Clauset and Nathan Eagle.
\newblock Persistence and periodicity in a dynamic proximity network.
\newblock {\em {DIMACS} Workshop on Computational Methods for Dynamic
  Interaction Networks}, 2007.

\bibitem{de2010inferring}
Munmun De~Choudhury, Winter~A. Mason, Jake~M. Hofman, and Duncan~J. Watts.
\newblock Inferring relevant social networks from interpersonal communication.
\newblock In {\em Proc. of the 19th Int. Conf. on World Wide Web}, pages
  301--310. ACM, 2010.

\bibitem{eagle2006reality}
Nathan Eagle and Alex Pentland.
\newblock Reality {M}ining: sensing complex social systems.
\newblock {\em Personal and Ubiquitous Computing}, 10(4):255--268, 2006.

\bibitem{erdos1959random}
P.~Erd{\H{o}}s and A.~R{\'e}nyi.
\newblock On random graphs {I.}
\newblock {\em Publ. Math. Debrecen}, 6:290--297, 1959.

\bibitem{freschi2009graph}
Valerio Freschi.
\newblock A graph-based semi-supervised algorithm for protein function
  prediction from interaction maps.
\newblock In {\em Learning and Intelligent Optimization}, pages 249--258.
  Springer, 2009.

\bibitem{gallagher2008using}
Brian Gallagher, Hanghang Tong, Tina Eliassi-Rad, and Christos Faloutsos.
\newblock Using ghost edges for classification in sparsely labeled networks.
\newblock In {\em Proc. of the 14th ACM SIGKDD Int. Conf. on Knowledge
  Discovery and Data Mining}, pages 256--264. ACM, 2008.

\bibitem{hansen2001model}
Mark~H. Hansen and Bin Yu.
\newblock Model selection and the principle of minimum description length.
\newblock {\em Journal of the American Statistical Association},
  96(454):746--774, 2001.

\bibitem{holme2012temporal}
Petter Holme and Jari Saram{\"a}ki.
\newblock Temporal networks.
\newblock {\em Physics Reports}, 519(3):97--125, 2012.

\bibitem{6137319}
Bing Hu, T.~Rakthanmanon, Yuan Hao, S.~Evans, S.~Lonardi, and E.~Keogh.
\newblock Discovering the intrinsic cardinality and dimensionality of time
  series using {MDL}.
\newblock In {\em IEEE 11th Int. Conf. on Data Mining (ICDM),2011}, pages
  1086--1091, Dec. 2011.

\bibitem{huang2005link}
Zan Huang, Xin Li, and Hsinchun Chen.
\newblock Link prediction approach to collaborative filtering.
\newblock In {\em Proc. of the 5th ACM/IEEE-CS Joint Conf. on Digital
  Libraries}, pages 141--142. ACM, 2005.

\bibitem{Liben-Nowell:2003}
David Liben-Nowell and Jon Kleinberg.
\newblock The link prediction problem for social networks.
\newblock In {\em Proc. of the 12th Int. Conf. on Information and Knowledge
  Management}, CIKM '03, pages 556--559, New York, NY, USA, 2003. ACM.

\bibitem{liu2007predicting}
Yan Liu and Zhenzhen Kou.
\newblock Predicting who rated what in large-scale datasets.
\newblock {\em ACM SIGKDD Explorations Newsletter}, 9(2):62--65, 2007.

\bibitem{DBLP:journals/corr/PeelC14}
Leto Peel and Aaron Clauset.
\newblock Detecting change points in the large-scale structure of evolving
  networks.
\newblock {\em CoRR}, abs/1403.0989, 2014.
\newblock Pre-print.

\bibitem{perra2012activity}
N.~Perra, B.~Gon{\c{c}}alves, R.~Pastor-Satorras, and A.~Vespignani.
\newblock Activity driven modeling of time varying networks.
\newblock {\em Scientific Reports}, 2, 2012.

\bibitem{ribeiro2013quantifying}
Bruno Ribeiro, Nicola Perra, and Andrea Baronchelli.
\newblock Quantifying the effect of temporal resolution on time-varying
  networks.
\newblock {\em Scientific Reports}, 3, 2013.

\bibitem{sarkar2005dynamic}
Purnamrita Sarkar and Andrew~W. Moore.
\newblock Dynamic social network analysis using latent space models.
\newblock {\em ACM SIGKDD Explorations Newsletter}, 7(2):31--40, 2005.

\bibitem{cambridge-haggle-2006-01-31}
James Scott, Richard Gass, Jon Crowcroft, Pan Hui, Christophe Diot, and
  Augustin Chaintreau.
\newblock {CRAWDAD} data set cambridge/haggle (v. 2006-01-31).
\newblock Downloaded from http://crawdad.org/cambridge/haggle/, Jan. 2006.

\bibitem{sulo2010meaningful}
Rajmonda Sulo, Tanya Berger-Wolf, and Robert Grossman.
\newblock Meaningful selection of temporal resolution for dynamic networks.
\newblock In {\em Proc. of the 8th Workshop on Mining and Learning with
  Graphs}, pages 127--136. ACM, 2010.

\bibitem{sun2007graphscope}
Jimeng Sun, Christos Faloutsos, Spiros Papadimitriou, and Philip~S. Yu.
\newblock Graphscope: Parameter-free mining of large time-evolving graphs.
\newblock In {\em Proc. of the 13th ACM SIGKDD Int. Conf. on Knowledge
  Discovery and Data Mining}, KDD '07, pages 687--696, New York, NY, USA, 2007.
  ACM.

\bibitem{Wagner:2008}
Neal Wagner and Zbigniew Michalewicz.
\newblock An analysis of adaptive windowing for time series forecasting in
  dynamic environments: Further tests of the {DyFor} {GP} {Model}.
\newblock In {\em Proc. of the 10th Conf. on Genetic and Evolutionary
  Computation}, GECCO '08, pages 1657--1664, New York, NY, USA, 2008. ACM.

\bibitem{zhu2002using}
Jianhan Zhu, Jun Hong, and John~G. Hughes.
\newblock Using markov chains for link prediction in adaptive web sites.
\newblock In {\em Soft-Ware 2002: Computing in an Imperfect World}, pages
  60--73. Springer, 2002.

\end{thebibliography}

\end{document}